\def\ptitle{A basis for variational calculations in {\mlarge d} dimensions}
\def\ptitler{A basis for variational calculations in $d$ dimensions}
\nopagenumbers
\hsize 6.0 true in 
\hoffset 0.25 true in 
\emergencystretch=0.6 in                 
\vfuzz 0.4 in                            
\hfuzz  0.4 in                           
\vglue 0.1true in
\mathsurround=2pt                        
\topskip=24pt                            
\def\nl{\noindent}                       
\def\np{\hfil\vfil\break}                
\def\ppl#1{{\leftskip=10cm\noindent #1\smallskip}} 
\def\title#1{\bigskip\noindent\bf #1 ~ \tr\smallskip} 
\font\tr=cmr10                          
\font\bf=cmbx10                         
\font\sl=cmsl10                         
\font\it=cmti10                         
\font\trbig=cmbx10 scaled 1500          
\font\tiny=cmr8                         
\font\mlarge=cmmi12 scaled 1200         
\def\mb#1{\hbox{\bf#1}}                 
\def\ng{>\kern -9pt|\kern 9pt}          
\def\hi#1#2{$#1$\kern -2pt-#2}          
\def\hy#1#2{#1-\kern -2pt$#2$}          

\def\half{{1 \over 2}}

\def\frac#1#2{{{#1}\over{#2}}}

\output={\shipout\vbox{\makeheadline
                                      \ifnum\the\pageno>1 {\hrule}  \fi 
                                      {\pagebody}   
                                      \makefootline}
                   \advancepageno}
\headline{\noindent {\ifnum\the\pageno>1 
                                   {\tiny \ptitler\hfil page~\the\pageno}\fi}}
\footline{}
\newcount\zz  \zz=0  
\newcount\q   
\newcount\qq    \qq=0  
\def\pref #1#2#3#4#5{\frenchspacing \global \advance \q by 1     
    \edef#1{\the\q}
       {\ifnum \zz=1 { %
         \item{[\the\q]} 
         {#2} {\bf #3},{ #4.}{~#5}\medskip} \fi}}
\def\bref #1#2#3#4#5{\frenchspacing \global \advance \q by 1     
    \edef#1{\the\q}
    {\ifnum \zz=1 { %
       \item{[\the\q]} 
       {#2}, {\it #3} {(#4).}{~#5}\medskip} \fi}}
\def\gref #1#2{\frenchspacing \global \advance \q by 1  
    \edef#1{\the\q}
    {\ifnum \zz=1 { %
       \item{[\the\q]} 
       {#2}\medskip} \fi}}
 \def\sref #1{~[#1]}
 
\def\references#1{\zz=#1
   \parskip=2pt plus 1pt   
   {\ifnum \zz=1 {\noindent \bf References \medskip} \fi} \q=\qq

\pref{\harr}{E. M. Harrell, Ann. Phys. (NY)}{105}{(1977)\ 379}{}
\pref{\detw}{L. C. Detwiler and J. R. Klauder, Phys. Rev. D\ }{11}{(1975)\ 1436}{}
\pref{\eks}{H. Ezawa, J. R. Klauder, and L. A. Shepp, J. Math. Phys.}{16}{(1975)\ 783}{}
\pref{\klac}{J. R. Klauder, Science}{199}{(1978)\ 735}{}

\pref{\za} {M. Znojil, J. Phys. A: Math. Gen. }{15}{2111 (1982)}{}
\pref{\zb} {M. Znojil, J. Phys. lett.}{101A}{66 (1984)}{}
\pref{\zc} {M. Znojil, J. Math. Phys. }{30}{23 (1989)}{}
\pref{\zd} {M. Znojil, J. Math. Phys. }{31}{108 (1990)}{}
\pref{\ze} {M. Znojil, Phys. Lett. A}{169}{415 (1992)}{}
\pref{\zf} {M. Znojil and P. G. L. Leach, J. Math. Phys.}{33}{2785 (1992)}{}
\pref{\zg} {M. Znojil, J. Math. Phys.}{34}{4914 (1993)}{}
\pref{\zh} {M. Znojil and R. Roychoudhury, Czech. J. Phys.}{48}{(1998)\ 1}{}
\pref{\zk} {M. Znojil, Phys. Lett. A}{255}{(1999)\ 1 }{}
\pref{\ks}{J. Killingbeck, J. Phys. A: Math. Gen. }{10}{L99 (1977)}{}
\pref{\kf}{J. Killingbeck, Phys. lett. A }{67}{ 13 (1978)}{}
\pref{\kf}{J. Killingbeck, Comp. Phys. Commun. }{18}{ 211 (1979)}{}
\pref{\ka}{J. Killingbeck, J. Phys. A: Math. Gen.}{13}{49 (1980)}{}
\pref{\kb}{J. Killingbeck, J. Phys. A: Math. Gen.}{13}{L231 (1980) }{}
\pref{\kd}{J. Killingbeck, J. Phys. A: Math. Gen.}{14}{ 1005 (1981)}{}
\pref{\kc}{J. Killingbeck, J. Phys. B: Mol. Phys.} {15}{829 (1982)}{}
\pref{\ke}{J. Killingbeck, G. Jolicard and A. Grosjean, J. Phys. A: Math. Gen.}{34}{L367 (2001)}{}
\pref{\ja}{M. J. Jamieson, J. Phys. B: At. Mol. Phys. }{16}{L391 (1983)}{}
\pref{\mill}{H. G. Miller, J. Math. Phys.}{35}{2229 (1994)}{}
\pref{\hajj}{F. J. Hajj, J. Phys. B: At. Mol. Phys. }{13}{4521 (1980)}{}
\pref{\kola}{H. J. Korsch and H. Laurent, J. Phys. B: At. Mol. Phys.}{14}{4213 (1981)}{}
\pref{\sol}{W. Solano-Torres, G. A. Est\'evez, F. M. Fern\'andez, and G. C. Groenenboom, J. Phys. A: Math. Gen.}{25}{3427 (1992)}{}

\pref{\bp}{E. Buendi\'a, F.J.G\'alvez, A. Puertas, J. Phys. A: Math. Gen. }{ 28}{6731 (1995)}{}
\pref{\ro}{A. K. Roy, Phys. lett. A}{ 321}{231 (2004)}{}
\pref{\pc}{Peace Chang and Chen-Shiung Hsue, Phys. Rev. A }{49}{4448 (1994)}{}

\pref{\rv}{R. K. Roychoudhury and Y P Varshni, J. Phys. A: Math. Gen. }{21}{3025 (1988)}{}

\pref{\aa}{V. C. Aguilera-Navarro, G.A. Est\'evez, and R. Guardiola, J. Math. Phys.}{31}{99 (1990)}{}
\pref{\ab}{V. C. Aguilera-Navarro and R. Guardiola, J. Math. Phys.}{32}{2135 (1991)}{}
\pref{\ac}{V. C. Aguilera-Navarro, F. M. Fern\'andez, R. Guardiola and J. Ros, J.Phys. A: Math. Gen}{25}{6379 (1992)}{}
\pref{\ad}{V. C. Aguilera-Navarro, A. L. Coelho and Nazakat Ullah, Phys. Rev. A}{49} {1477 (1994)}{}
\pref{\af}{V. C. Aguilera-Navarro and R. Guardiola, J. Math. Phys.}{32}{2135 (1991)}{}
\pref{\ag}{V. C. Aguilera-Navarro and Ley Koo, Int. J. Theo. Phys.} {36} {157-1666} {(1997)}{}
\pref{\ah}{R. Guardiola and J. Rose, J. Phys. A: Math. Gen. }{25}{1351 (1992)}{}
\pref{\ha}{R. L. Hall and N. Saad, Can. J. Phys.}{73} {493} {(1995)}{}
\pref{\hb}{R. L. Hall and N. Saad, J. Phys. A: Math. Gen. }{29}{ 2127 (1996)}{}
\pref{\hc}{R. L. Hall, N. Saad, and A. von Keviczky, J. Math. Phys. }{39}{6345 (1998)}{}
\pref{\hd}{R. L. Hall and N. Saad, J.Phys. A: Math. Gen. }{33}{5531 (2000)}{}
\pref{\he}{R. L. Hall and N. Saad, J.Phys. A: Math. Gen. }{34}{1169 (2001)}{}
\pref{\hf}{R. L. Hall, N. Saad and A. von Keviczky, J. Phys. A }{34}{1169 (2001)}{}
\pref{\hg}{R. L. Hall, N. Saad, and A. von Kevicsky, J. Math. Phys. }{43}{94(2002)}{}
\pref{\hh}{N. Saad and R. L. Hall, J. Phys. A.: Math. Gen. }{35}{4105 (2002)}{}
\pref{\hi}{R. L Hall and N. Saad, J. Math. Phys. } {43} {94 (2002)}{}
\pref{\hl}{N. Saad, R.L. Hall, and A. von Keviczky, J. Math. Phys. }{44}{ 5021 (2003)}{} 
\pref{\hm}{N. Saad, R. L. Hall, and A. von Keviczky, J. Phys. A: Math. Gen. }{36}{487 (2003)}{}

\pref{\la}{M. Landtman, phys. lett. A }{175}{335 (1993}{}
\pref{\bg}{E. Buendi\'{i}a, F.J.G\'alvez, A. Puertas, J. Phys. A: Math. Gen. }{28}{6731 (1995)}{}
\pref{\va}{Y. Varshni, phys. lett. A }{183}{9 (1993)}{}
\pref{\sz}{Shi-Hai Dong and Zhong-Qi Ma, J. Phys. A: Math. Gen. }{31}{9855 (1998)}{}
\pref{\fer}{F. M. Fern\'andez, Phys. Lett. A}{160}{511 (1991)}{}
\pref{\mdel}{M. de Llano, Rev. Mex. Fis.}{27}{(1981)\ 243 }{}
\pref{\flyn}{M. F. Flynn, R. Guardiola, and M. Znojil, Czech. J. Phys.}{41}{(1993)\ 1019}{}
\pref{\nag}{N. Nag and R. Roychoudhury, Czech. J. Phys.}{46}{(1996)\ 343}{} 
\pref{\esta}{E. S. Est\'evez-Bret\'on and G. A. Est\'evez-Bret\'on,  J. Math. Phys.}{34}{(1993)\ 437}{}
\pref{\moa}{O. Mustafa and M. Odeh, J. Phys. B}{32}{3055 (1999)}{}
\pref{\mob}{O. Mustafa and M. Odeh, J. Phys. A}{33}{5207 (2000)}{}
\pref{\hau}{A P Hautot, J. Math. Phys. }{13}{ 710 (1972)}{}
\pref{\krt}{P. Kumar, M. Rusiz-Altaba, and B. S. Thomas, Phys. Rev. Lett. }{57}{2759 (1986)}{}
\pref{\weu}{Wai-Yee Keung, Eve Kovacs, and Uday P. Sukhatme,  Phys. Rev. Lett.}{60}{41 (1988)}{}
\bref{\movr}{H. A. Mavromatis}{Exercises in Quantum Mechanics}{Kluwer, Dordrecht, 1991}{}
\bref{\som}{A. Sommerfeld}{Partial Differential Equations in Physics}{Academic, New York, 1949}{The Laplacian in $N$ dimensions is discussed on pp. 227, 231}
\pref{\loew}{P. -O. L\"owdin, J. Chem. Phys.}{18}{365 (1950)}{}
\pref{\davies}{P. I. Davies, N. J. Higham, and F. Tisseur, Siam J. Matrix Anal. Apps.}
{23}{472 (2001)}{}
\bref{\reed}{M. Reed and B. Simon}{Methods of Modern Mathematical Physics IV: Analysis of Operators}{Academic, New York, 1978}{The min-max principle for the discrete spectrum is discussed on p75}
\pref{\kp}{R. S. Kaushal and D Parashar, Phys. Lett. A }{170}{335 (1992)}{} 
\bref{\roy}{D. K. Roy}{Quantum Mechanics Tunneling and its applications}{World Scientific, Singapore, 1986}{; L. A. MacColl, Phys. Rev. {\bf 40}, 261 (1932)}
\pref{\den} {D. M. Dennison and G. E. Uhlenbeck, Phys. Rev. }{41}{261 (1932)}{; F. T. Wall and G. Glockler, J. Chem. Phys. {\bf 5}, 314 (1937).}
\pref{\bd}{G. R. P. Broges, A. de Souza Dutra, Elso Drigo, and J. R. Ruggiero, Can. J. phys. }{81}{1283 (2003)}{} 
\bref{\gj}{G. Harvey and J. Tobochnik}{An introduction to computer simulation methods: applications to physical systems, 2nd ed.}{Addison-Wesley, Reading, Mass. 1996}{p. 631}
 }
 \references{0}    

\ppl{CUQM-105}\ppl{math-ph/0410035} 
\ppl{October 2004}\medskip 

\tr 
\vskip 1.0true in
\centerline{\trbig \ptitle}
\vskip 0.5 true in
\baselineskip 12 true pt 
\centerline{\bf Richard L. Hall$^1$, Qutaibeh D. Katatbeh$^2$, and Nasser Saad$^3$}
\vskip 0.2 true in
\centerline{\sl $^{(1)}$ Department of Mathematics and Statistics,}
\centerline{\sl Concordia University,}
\centerline{\sl 1455 de Maisonneuve Boulevard West,}
\centerline{\sl Montr\'eal, Qu\'ebec, Canada H3G 1M8.}
\vskip 0.2 true in
\centerline{\sl $^{(2)}$ Department of Mathematics and Statistics,}
\centerline{\sl Faculty of Science and Arts,}
\centerline{\sl Jordan University of Science and Technology,}
\centerline{\sl Irbid 22110, Jordan.}
\vskip 0.2 true in
\centerline{\sl $^{(3)}$ Department of Mathematics and Statistics,}
\centerline{\sl University of Prince Edward Island,}
\centerline{\sl 550 University Avenue, Charlottetown,}
\centerline{\sl PEI, Canada C1A 4P3.}\vskip 0.2 true in
\bigskip\bigskip
\baselineskip = 18true pt  
\centerline{\bf Abstract}\medskip
In this paper we derive expressions for matrix elements $(\phi_i,H\phi_j)$ for the Hamiltonian $H=-\Delta+\sum_q a(q)r^q$ in $d\geq 2$ dimensions.  The basis functions in each angular momentum subspace are of the form $\phi_i(r)=r^{i+1+(t-d)/2}e^{-r^p/2},\ i\geq 0,\ p > 0, t > 0.$ The matrix elements are given in terms of the Gamma function for all $d$.  The significance of the parameters $t$ and $p$ and scale $s$ are discussed.  Applications to a variety of potentials are presented, including potentials with singular repulsive terms of the form $\beta/r^\alpha$,~ $\alpha,\beta > 0,$ perturbed Coulomb potentials $-D/r + B r + Ar^2,$ and potentials with weak repulsive terms, such as $-\gamma r^2 + r^4,$ $\gamma > 0.$
  
\medskip\noindent PACS~~03.65.Ge,~31.15.Bs,~02.30.Mv.
\np
  \title{1.~~Introduction}
We study quantum mechanical Hamiltonians $H = -\Delta + V(r)$ in $d\geq 2$ dimensions, where $V$ is a spherically-symmetric potential that supports discrete eigenvalues, $r = |\mb{r}|,$ $\mb{r}\in \Re^d.$ We estimate the spectrum of $H$ in an $n$-{dimensional} trial space lying inside an angular-momentum subspace labelled by $\ell$ and spanned by radial functions with the form
$$\phi(r) = \sum_{i = 0}^{n-1} c_i r^{i+1+(t-d)/2}e^{-\half r^p},\quad t > 0.\eqno{(1.1)}$$
\nl If the potential is chosen to be a linear combination of powers
$$V(r) = \sum_{q}a(q)r^q,\eqno{(1.2)}$$
\nl then all the matrix elements of $H$ may be expressed explicitly in terms of the Gamma function.  The expressions obtained will be functions of the parameters $t$ and $p$, and also of a scale parameter $s$ to be introduced later. If the potential $V$ is highly singular, the parameter $t$ must be chosen sufficiently large so that $\langle V\rangle$ exists. The advantages of the particular form chosen for the radial functions will become clear in the development. Thus we have $n + 3$ variational parameters with which to optimize upper estimates to the spectrum of $H$, with one degree of freedom being employed for normalization. 
 
Systems with Hamiltonians of this type have enjoyed wide attention in the literature of quantum mechanics\sref{\harr-\weu}. This interest arises particularly from the usefullness of these  problems as models in atomic and molecular physics.  Many numerical and analytical techniques have been used to tackle Hamiltonians of this form.  In Section~2 we derive general matrix elements and show how the minimization with respect to scale $s$ can be easily included. In section~3 we discuss some numerical issues not the least of which is the usefulness of the reduction of the matrix eigen equations to symmetric form by first diagonalizing the `normalization' matrix $N = [(\phi_i,\phi_j)].$  The dependence of the eigenvalues on the parameters $\{p,t,s\}$ may be rather complicated.  Since changes to scale $s$ do not involve the recomputation of the basic matrix elements, a policy which emerges is to fix $n$, always optimize fully with respect to scale $s,$ and, if necessary, optimize approximately with respect to $t$ and $p$by exploring a few values; if higher accuracy is required, a full optimization is undertaken, or $n$ is increased. In Section~4 the matrix elements are applied to a variety of problems and the results are compared with those found in earlier work.   
 We suppose that the  Hamiltonian operators in this paper have domains ${\cal D}(H)\subset L^2(\Re^d),$ they are bounded below, essentially self adjoint, and have at least one discrete eigenvalue at the bottom of the spectrum.  This, of course, implies that the potential cannot be dominated by repulsive terms. Because the potentials are spherically symmetric, the discrete eigenvalues $E_{n\ell}^d$ can be labelled by two quantum numbers, the total angular momentum $\ell = 0,1,2,\dots,$ and a `radial' quantum number, $n = 0,1,2,\dots,$ which counts the eigenvalues in each angular-momentum subspace. These eigenvalues satisfy the relation $E^d_{n\ell}\le E^d_{m\ell},\ n<m.$ With our labelling convention, the eigenvalue $E^d_{n\ell}(q)$ in $d\geq 2$ spatial dimensions has degeneracy $1$ for $\ell=0$ and, for $\ell>0,$ the degeneracy is given\sref{\movr} by the function $\Lambda(d,\ell)$, where
$$\Lambda(d,\ell)=(2\ell+d-2)(\ell+d-3)!/\{\ell!(d-2)!\},\quad d \geq 2, \ \ell>0.\eqno{(1.3)} $$
Many techniques have been applied to approximate the spectrum of singular potentials of the form (1.2) using perturbation, variational, and geometrical approximation techniques\sref{\harr-\weu}. Exact solutions for the energy may be obtained in some special cases by first choosing a wave function with parameters, and then finding a potential of the form (1.2) for which this wave function is an eigenfunction; this is possible only when certain constraints are satisfied between the parameters $\{a(q)\}$, as we shall discuss later. 
  \title{2.~~Matrix elements}
We consider first the action of the Laplacian in $d$ dimensions on a wave function $\Psi(\mb{r}) = \psi(r) Y_{\ell}(\theta_0,\theta_1,\dots,\theta_{d-1})$ with a spherically-symmetric factor $\psi(r)$ and a generalized spherical harmonic factor $Y_{\ell}.$  If we remove the spherical harmonic factor after the action of the Laplacian on $\Psi$ we obtain\sref{\som} 
$$\frac{\Delta \Psi} {Y_{\ell}} = \psi''(r) +\frac{d-1}{r}\psi'(r) - \frac{\ell(\ell+d-2)}{r^2}\psi(r).\eqno{(2.1)}$$
The radial Schr\"odinger equation for a spherically symmetric potential $V(r)$ in $d$-dimensional space is therefore given by
$$-{d^2\psi\over dr^2}-{d-1\over r}{d\psi\over dr}+{l(l+d-2)\over r^2}\psi+V(r)\psi=E\psi,\quad \psi(r)\in L^2([0,\infty),r^{d-1}dr)\eqno(2.2)$$ 
\nl A correspondence to a problem on the half line in one dimension with a Dirichlet boundary condition at $r = 0$ is obtained with the aid of a radial wave-function $R(r)$ defined by
$$R(r)=r^{(d-1)/2}\psi(r),\quad d\geq 2,~~R(0) = 0.\eqno{(2.3)}$$
\nl If we now re-write (2.2) in terms of this new radial function, we obtain the following Schr\"odinger equation for a problem on the half line 
$$HR = -{d^2R\over {dr^2}}+ UR=ER,\quad R\in L^2([0,\infty),dr),\eqno(2.4)$$
\nl where the effective potential $U(r)$ is given by
$$U(r) = V(r) + {{(2\ell+d-1)(2\ell+d -3)}\over {4  r^2}},\eqno{(2.5)}$$
\nl and $H = -{{d^2}\over {dr^2}} + U$ is the effective Hamiltonian. We note that in (2.5), $d$ and $l$ enter into the equation only in the combination $2l+d$ in $U(r)$: consequently, the solutions for a given central potential $V(r)$ are the same provided $d+2l$ remains unaltered.  In this setting, our trial wave functions now have the explicit form 
$$R_i(r)=r^{(t+1)/2+i}\exp(-r^p/2) \in L^2([0,\infty),dr).\eqno{(2.6)}$$
\nl Thus we have for the general radial function in our trial space
$$R(r) = \sum_{i = 0}^{d-1}c_i R_i(r).\eqno{(2.7)}$$ 
\nl The matrix elements we seek (in a given angular momentum subspace) are given by
$$H_{ij}=\left(R_i,-R_j^{\prime\prime}\right)+ \sum_q a(q)\left(R_i, r^q R_j\right).\eqno{(2.8)} $$ 
\nl For each potential term $r^q,$ if we everywhere omit the constant angular factor (equal to $4\pi$ in the case $d = 3$), we find the following fomulae, expressed now in terms of the $L^{2}([0,\infty), dr)$ inner product:
$$\eqalign{P_{ij}(q,p,t) &= \left(R_i,r^q R_j\right) = \int_0^{\infty}r^{i+j+1+t+q}e^{-r^p}dr,\quad i,j = 0,1,2,\dots,\cr & = \frac{1}{p}\Gamma\left(\frac{i+j+t+q+2}{p}\right),\quad t > -(q + 2).}\eqno{(2.9)}$$
\nl This type of integral is found by setting $x = r^p,$ and using the differential relation $r^kdr = (1/p)x^{(k+1-p)/p} dx$  and the definition of the Gamma function.  The normalization integrals are special cases of (2.9), namely
$$N_{ij}(p,t) = \left(R_i,R_j\right) = P_{ij}(0,p,t)= \frac{1}{p}\Gamma\left(\frac{i+j+t+2}{p}\right).\eqno{(2.10)}$$
\nl After some algebraic simplifications we find that the corresponding kinetic energy matrix elements $K_{ij}(p,t) = -\left(R_i,R_j^{\prime\prime}\right)$ are given by
$$K_{ij}(p,t) = {1\over {4p}}\Gamma\left({{i+j+t}\over{p}}\right)
\left[(2\ell+d-1)(2\ell+d-3) +1 -(i-j)^2 + p(i+j+t)\right],\quad t > 0.\eqno{(2.11)}$$
\nl We note that these terms of the Hamiltonian matrix elements $H_{ij}$ are all symmetric under the permutation $(i j)$ (because of Hermiticity), and invariant with respect to changes in $d$ and $\ell$ that leave the form $2\ell+d$ invariant. These formulae may be used as they stand for all dimensions $d \geq 2$ provided that $t>0$ is chosen sufficiently large $t > -(2+\hat{q})$ to control the most singular potential term $r^{\hat{q}}.$  We note, in addition, that the choice $\{d = 3,~\ell = 0\}$ also provides the odd-parity solutions in one dimension.  

\medskip 
   
We now consider the problem of minimizing $(R,HR)$ with respect to the vector $v$ of coefficient $\{ c_i\}_{i=0}^{n-1}$ subject to the constraint that $(R,R)=1$. We immediately obtain the necessary condition: 
$$Hv= {\cal E} N v \eqno{(2.12)}$$
By the min-max characterization of the spectrum\sref{\reed}, the eigenvalues of this matrix equation are upper bounds to the unknown exact eigenvalues $E_{i\ell}, i=0,1,2,\dots n-1.$ We assume that these discrete eigenvalues of the underlying operator $H$ are either known to exist, or indeed are demonstrated to exist by the results of this variational estimate.
\nl By considering scaled radial wave functions of the form 
$$R_s(r)=R(r/s),\eqno{(2.13)} $$ 
we find that factors of $s$ remain only according to the dimensions of the terms. In effect, when using the scaled wave functions (2.13), we can leave the matrix $N$ unchanged and replace the matrix for $H$ by
$$H_{ij}(s)={1\over s^2}K_{ij}(p,t)+\sum_q a(q)s^q P_{ij}(q,p,t).\eqno{(2.14)} $$
Thus the upper bounds we seek are provided by the eigenvalues of the matrix equation
$$H(s)v = {\cal E} N v,\eqno{(2.15)}$$
 which now depend, for a given $n$ and $\ell,$ on $s, p$ and $t$ and we write
$$E_{i\ell} \leq {\cal E}_{i\ell}={\cal E}_{i\ell}(p,t,s),\quad i = 0,1,2 \dots n-1.\eqno(2.17)$$
\nl The problem now is to find these upper estimates and minimize them with respect to the three parameters $\{p,t,s\}$.
\title{3.~~Some numerical considerations}
Rather than solving the general matrix eigenequation (2.7) directly, it is often desirable to use the fact that $N$ is positive definite to transform the problem to symmetric form.  In physics literature this is sometimes called a L\"owdin transformation\sref{\loew} and is equivalent analytically to converting the basis functions to an orthonormal set by applying the Gram-Schmidt procedure. We first diagonalize $N$ with the aid of an orthogonal matrix, say $S.$ We then get $S^TNS=M^{-2}:$ the square root $M$ exists because $N$ positive definite, which implies that the diagonal matrix has only positive eigenvalues. The original problem $(2.12)$ (or the scaled version (2.15)) may now be written as
$$Hv=\lambda Nv\rightarrow S^THSS^Tv=\lambda S^TNSS^Tv=\lambda M^{-2}S^Tv \eqno{(3.1)}$$
If we multiply on the left by symmetric diagonal matrix $M$ we obtain
$$MS^THSMM^{-1}S^Tv=\lambda M^{-1}S^Tv\eqno{(3.2)} $$
If we now write ${\cal {H}}=MS^THSM,$ and $u=M^{-1}S^Tv,$ we obtain the reduction 
$${\cal {H}}u=\lambda u,\eqno{(3.3)}$$
\nl where ${\cal{H}}^T={\cal H}.$ This is the symmetric alternative to our original eigenvalue problem.  It has also been shown that the Cholesky decomposition\sref{\davies} in which the matrix $N$ is written $N = L^TL,$ where $L$ is upper triangular, is often numerically faster and more stable than finding the square root $M$.  Computer algebra systems often allow one to solve these problems directly without knowing which method is in fact implemented; the main purpose of our remarks is to show constructively that solutions are always possible.\medskip
Another issue is to do with the Gamma function generating large numbers before (or without) the symmetrization of $H$.  To deal with this problem we have found it useful at an early stage to divide all the matrix elements by $(N_{ii}N_{jj})^\frac{1}{2}.$\medskip
Ideally the matrix eigenvalues should simply be optimized with respect to the parameters $\{p,t,s\}.$  In practice this is not always a trivially easy task.  Typically, one chooses the basis dimension $n$ and the angular momentum $\ell,$ and then finds the $n$ eigenvalues.  These numbers must be sorted to find, say, the $k$th eigenvalue ${\cal E}_{k\ell}(p,t,s),$ and finally this function must be optimized with respect to the three parameters. This appears to be straightforward until one realizes that the matrix eigenvalue problem must be re-solved for each choice of the parameters and, of course, the original ordering can be upset.  Logically the $k$th always has the same numerical meaning but the effect is to make the function ${\cal E}_{k\ell}(p,t,s)$ complicated. It is helpful to note that the basic matrices $N(p,t),$ $P(q,p,t)$ and $K(p,t)$ do not depend on $s:$ the Hamiltonian matrix $H$ depends on $s$ by the scaling equation (2.10). In order to reduce the difficulty of the search for a minimum we have sometimes found it useful to fix $p$ and $t$ and to minimize at first only with respect to $s;$ if necessary a graph can be plotted of the dish-shaped function ${\cal E}_{k\ell}(s)$ to give a picture of the minimum.  This task may then be repeated for some other choices of $p$ and $t$.  In many cases an algorithm such as Nelder-Mead tackles the full minimization problem very effectively and there is no more ado concerning it.  We shall make some comments concerning these matters along with the applications described in section~4 below.      
  \title{4.~~Applications}
We may immediately employ the matrix elements found to solve the eigenvalue problems for
 the general family of Hamiltonians given by
$$H=-\Delta + \sum_q a(q) r^q,\eqno(4.1) $$
One family we shall study in particular is the class of anharmonic singular Hamiltonians
$$H=-{d^2\over dr^2}+r^2+\sum\limits_{q=0}^N {{\lambda_q}\over{r^{\alpha_q}}},\quad r\in[0,\infty)\eqno(4.2)$$
where $\alpha_q$ and $\lambda_q$ are positive real numbers, and we assume that the exact wave function $\psi$ of $H$ satisfies a {\it Dirichlet boundary condition}, namely $\psi(0) = 0$. Inverse power-law potentials $V(r)=\sum\limits_{q=0}^N {{\lambda_q}/r^{\alpha_q}}$ appear in many areas of physics and for this reason have been widely investigated. The spiked harmonic oscillator Hamiltonian, for example,
$$H(\alpha,\lambda)=-{d^2\over dr^2}+r^2+{\lambda\over{r^{\alpha}}},\quad \alpha>0, \lambda>0\eqno(4.3)$$
has been the subject of many mathematical studies which have greatly improved the understanding of singular perturbation theory\sref{\harr,\hl}.
Many different methods\sref{\harr-\weu} have been used to study the anharmonic singular Hamiltonians (4.2), such as numerical integration of the differential equation, perturbative schemes specifically developed for this class of Hamiltonian, and variational methods. Among the various methods, the variational method is widely used for calculating energies and wave functions since it has the advantage that the eigenvalue approximations are upper bounds\sref{\reed}. Many variational techniques used in the literature were design to solve specific classes of Hamiltonian such as (4.3). Aguilera-Navarro et al\sref{\aa}, for example, reported a variational study for the ground-state energy of the spiked harmonic oscillator (4.3) valid only for $\alpha<3$. Their study makes use of the function space spanned by the exact solutions of the Schr\"odinger equation for the linear harmonic oscillator Hamiltonian, supplemented by a Dirichlet boundary condition $\psi(0)=0$, namely,
$\psi_n(r)=A_n e^{-r^2/2}H_{2n+1}(r),\ A_n^{-2}=4^n(2n+1)!\sqrt{\pi},\ n=0,1,2,\dots$
where $H_{2n+1}(r)$ are the Hermite polynomials of odd degree. The matrix elements of the operator $r^{-\alpha}, \alpha< 3$, in this orthonormal basis were given as
$$
r_{mn}^{-\alpha}=(-1)^{m+n}{\sqrt{(2m+1)!(2n+1)!}\over 2^{m+n}~m!~n!}{\Gamma({3\over 2})\over \Gamma(n+{3\over 2})}\sum\limits_{k=0}^m (-1)^k\pmatrix{m\cr k\cr}{\Gamma(k+{3-\alpha\over 2})\Gamma(n+{\alpha\over 2}-k)\over 
\Gamma(k+{3\over 2})\Gamma({\alpha\over 2}-k)},\quad \alpha<3.
$$
A variational analysis was carried out and the ground-state upper bounds were reported for the case of $\alpha = {5/ 2}$. Fernandez\sref{\fer}, soon afterwards, design a particular trial function $\psi(r)=r^{k+1} e^{-{1\over 2} sr^{-2}-{1\over 2}tr^2}, s\geq 0, t>0,$  to study the ground-state energy of (4.3) for $\alpha$ even integer and for arbitrary value of $\lambda>0$. An upper bound to the ground-state of (4.2) was found by a minimization with respect to $\{s,t\}$ of 
$E_0(s,t)=((1-t^2)I_4+3tI_2+2stI_{-2}-s^2I_{-4}+\lambda I_{2-\alpha})/I_2$
where $I_n(s,t)=\int_{0}^\infty r^n \exp(-s/r^2-tr^2)dt, -\infty<n<\infty$. These variational results however were not very accurate, even for arbitrary large value of $\lambda$ owing to the accumulated error in the computation of $I_n(s,t)$. An interesting consequence of Fernandez's work was, however, the exact solution of very particular class of (4.3), namely $H=-{d^2/dr^2}+r^2+{9/64} r^{-6}$, where the exact wavefunction in this case reads $\psi(r)=r^{3\over 2} e^{-{3\over 16}r^{-2}-{1\over 2}r^2}$ and the exact ground-state energy is $E_0=4$. Aguilera-Navarro et al\sref{\ad} afterwards designed another trial function particularly devoted to analyze the ground-state energy of the Hamiltonian $H(4,\lambda)$. Non-orthogonal basis set of trial wave functions were introduced by means of
$\psi_n(r)=A_n~{}_1F_1(-n;{3\over 2};r^2)\exp(-ar^2-{b\over r}),\ n=0,1,2,\dots$
where $A_n$ is the normalization constant and ${}_1F_1(-n;{3/2};r^2)$ is the confluent hypergeometric function. The expressions for the matrix elements $H_{mn}(4,\lambda)$ were given by
$$H_{mn}=\sum\limits_{q=0}^n\sum\limits_{q=0}^m {(-n)_{p}(-m)_q\over ({3\over 2})_p({3\over 2})_q}{A_mA_n\over p!~q!}[(4m+3)I(2p+2q+4)+2\sqrt{\lambda}I(2p+2q+1)-4q\sqrt{\lambda}(2p+2q-1)]$$
where the definite integrals $I(u)=\int_0^\infty r^u \exp(-r^2-(2\sqrt{\lambda}/r))dr$ were computed by means of the recursive relations $(u+1)I(u)=(u-1)I(u-2)+2\sqrt{\lambda}I(u-3)$. The shifted factorial $(a)_n$ is defined by
$$(a)_0=1,\quad (a)_n=a(a+1)(a+2)\dots (a+n-1), \quad{\rm for}\ n = 1,2,3,\dots,\eqno(4.4)$$
which may be expressed in terms of the Gamma function by $(a)_n={\Gamma(a+n)/ \Gamma(a),}$ when $a$ is not a negative integer $-m$, and, in these exceptional cases, $(-m)_n = 0$ if $n > m$ and otherwise $(-m)_n = (-1)^n m!/(m-n)!.$
The ground-state of (4.3) with $\alpha = 4$ then follows by diagonalization of $H$ in the nonorthogonal basis. This particular study was then extended\sref{\ag} to provide a global analysis of the ground and excited states for the successive values of the orbital angular momentum of the super-singular plus quadratic potential $r^2 +{\lambda/r^4}$. Another variational study of the ground state of (4.2) was introduced by Hall {\it et al}\sref{\ha} where three parameters trial functions 
$\psi(r)=r^{p+\epsilon}\exp(-\beta r^q),\ p={(\alpha-1)/2}$ were used to approximate upper bounds of the ground-state of (4.3) for arbitrary $\alpha$ and $\lambda$ through the minimization of the right-side of the inequality $E_0\leq E_0^U$, where
$$E_0^U=\min\limits_{\epsilon,\beta,q> 0}\bigg[{q\over 2}(2\beta)^{2/q}\bigg[(2p+q+2\epsilon-1)g_1-{2\over q}(p+\epsilon)(p+\epsilon -1)g_2-{q\over 2}g_3\bigg]+\bigg({1\over 2\beta}\bigg)^{2/q} g_4+\lambda (2\beta)^{\alpha/q}g_5\bigg]/g_6$$
and
$$\matrix{g_1=\Gamma\bigg({2p+2\epsilon+q-1\over q}\bigg),& g_2=\Gamma\bigg({2p+2\epsilon-1\over q}\bigg),& g_3=\Gamma\bigg({2p+2\epsilon+2q-1\over q}\bigg)\cr
g_4=\Gamma\bigg({2p+2\epsilon+3\over q}\bigg),& g_5=\Gamma\bigg({2p+2\epsilon-\alpha+1\over q}\bigg),& g_6=\Gamma\bigg({2p+2\epsilon+1\over q}\bigg).\cr}
$$
In attempt to provide a comprehensive variational treatment of the spiked harmonic oscillator Hamiltonian (4.3), for ground-state energy as well for excited states, independent of particular choices of the parameters  $\alpha$ and $\lambda$, Hall et al \sref{\hc-\hm} based their variational analysis of the singular Hamiltonian (4.1) on an exact soluble model which itself has a singular potential term. They have suggested and used trial wave functions constructed by means of the superposition of the orthonormal functions of the exact solutions of  the Gol'dman and Krivchenkov Hamiltonian 
$$H_0=-{d^2\over dr^2}+r^2+{A\over r^2}.\eqno(4.5)$$
The Hamiltonian is the generalization of the familiar harmonic oscillator in 3-dimension $ -{d^2/dr^2}+r^2+{l(l+1)/r^2}$ where the generalization lies in the parameter $A$ ranging over [$0,\infty)$ instead of only values determined by the angular momentum quantum numbers $l=0,1,2,\dots$. The energy spectrum of the Schr\"odinger Hamiltonian $H_0$ is given, in terms of parameter $A$ as $${E}_n=2(2n+\gamma),\quad n=0,1,2,\dots,\eqno(4.6)$$
in which $\gamma=1+\sqrt{A+{1\over 4}}$ and the normalized wavefunctions are
$$
\psi_n(r)=(-1)^n\sqrt{{2(\gamma)_n}\over n! \Gamma(\gamma)} r^{\gamma-{1\over 2}} e^{-{1\over 
2} r^2}{}_1F_1(-n;\gamma;r^2).\eqno(4.7)
$$ 
Here ${}_1F_1$ is the confluent hypergeometric function
$${}_1F_1(-n;b;z)=\sum\limits_{k=0}^n {{(-n)_kz^k}\over {(b)_kk!}},\quad\hbox{($n$-degree polynomial in $z$)}.\eqno(4.8)$$
Explicit matrix elements of the Hamiltonian (4.3) can often be found in this orthonormal basis.
 For instance, the matrix elements of the singular operator $\lambda r^{-\alpha}$ assume the form
$$
r_{mn}^{-\alpha}=(-1)^{n+m}{({\alpha\over 2})_n\over
(\gamma)_n}{{\Gamma(\gamma-{\alpha\over 2})}\over
\Gamma(\gamma)}\sqrt{{(\gamma)_n(\gamma)_m}\over {n!m!}}
{}_3F_2\bigg(\matrix{-m,{\gamma-{\alpha\over 2}},{1-{\alpha\over 2}}\cr
	\gamma,1-{\alpha\over 2}-n\cr}\bigg|1\bigg),\eqno(4.9)
$$
where the hypergeometric function ${}_3F_2$
is defined by
$${}_3F_2\bigg(\matrix{-m,a,b\cr
	c,d\cr}\bigg|1\bigg)=\sum\limits_{k=0}^m{(-m)_k(a)_k(b)_k\over (c)_k(d)_k\ k!},\quad (m-\hbox{degree polynomial)}.$$
Upper bounds to the energy levels of the Hamiltonian (4.3) then follow by diagonalization of $H$ in the orthonormal basis (4.5). In the case where $\alpha$ is a non-negative even number $\alpha=2,4,6,\dots$, the hypergeometric function ${}_3F_2$ in (4.9) can be regarded as a polynomial of degree ${\alpha\over 2}-1$ instead of an $m$-degree polynomial. Consequently the matrix elements assumes much simpler expressions which are useful in numerical computational. For $\alpha \neq 2,4,6,\dots$, the variational computational were then based on direct use of the matrix elements in terms of the  hypergeometric function ${}_3F_2$. According to our discussion up to this point, it is clear that most of the variational methods developed in the literature were specifically design to solve the eigenvalue problem of different classes of the singular Hamiltonian (4.2). No basis set or trial wave function were design to treat a problem such as the singular potentials which at the same time can be used, say, for Hamiltonians with polynomial type potentials. The purpose of our basis introduced in section (2) and (3) is to have avaliable at our disposal a working variational approach that can be used without a particular references to specific potentials or special values for the parameters involve.   In the next we apply the matrix elements discussed in section (2) and (3) to solve a number of different eigenvalue problems.
\np
\title{4.1~~Spiked Harmonic Oscillators}
We start our applications by investigating the energy levels of the spiked Harmonic Oscillator Hamiltonian (4.3). As we mentioned in section 3, the problem of finding the eigenvalues reduces to diagnalizing the real symmetric matrix ${\cal {H}}=MS^THSM$. For $\alpha=2$, the Hamiltonian (4.3) admits an exact solutions (4.6). Thus it serves as a benchmark for our variational approach. In Table (1), we report our upper bounds for the ground-state of the spiked harmonic oscillator Hamiltonian (4.3) for several values of the parameters $\lambda$ and $\alpha={1\over 2},1,{3\over 2},2,{5\over 2}$ along with some results obtained in the literature. For $\alpha=2$, with $m=n=0$, Table 1 shows that minimization over the three variables $\{p,t,s\}$ yields excellent agreement with the exact solutions (4.6). Such results can be explained by observing that direct substitution of the trial wave function $\psi_0(r)$ into the eigenvalue problem
$$H\psi_0=-{d^2\psi_0\over dr^2}+(r^2+{\lambda\over r^\alpha})\psi_0=E_0\psi_0\eqno{(4.10)}$$
yields for  $r\rightarrow 0$ that
$${{t^2-1}\over { 4r^2}}+{p^2\over 4r^{2-2p}}-{(t+1)p\over 2 r^{2-p}}-{\lambda\over r^\alpha}=0\eqno{(4.11)}$$
Consequently, for $\alpha=2$ and $p>0$, the value $t=1+|1-\sqrt{1+4\lambda}|$ yields the best possible value of $t$.  As for $\alpha < 2$, similar reasoning yields for $r\rightarrow 0$ that $t=1$ is an excellent {\it initial} approximation for $t,$ that is to say, suitable for starting the minimization process.  In Table 2, we present a comparison between different variational approaches for computing upper bounds to the ground-state of the Hamiltonian $H=-{d^2/dr^2}+r^2+\lambda r^{-5/2}$ for $\lambda>0$, where the diagonalization of ${\cal {H}}$, Eq.(3.3), was carried out in variational spaces of different dimensions $n$. In Table 3, we report our variational computation for upper bounds to the ground-state energy of the Hamiltonian $H=-{d^2/dr^2}+r^2+\lambda r^{-4}$ along with the eigenvalues reported in the literature. In Table 4, we extended our variational analysis to study the Hamiltonian $H=-{d^2/dr^2}+r^2+l(l+1)/r^2+\lambda/r^{4}$ for $\lambda\ll 1$ and for several values of $l$. We compare our results with the those in the literature, along with `exact' eigenvalues obtained by direct numerical computation of the corresponding Schr\"odinger equation. In order to keep the number of tables of results to a minimum, we first mention the case of $V(r)=r^2+{9\over 64}{1\over r^6}$ which yields the exact energy $E = 4$: by using our variational approach we obtain an upper bound of $E=4.0000006$ for $n=15$ with $p= 0.73, t =7.09,$ and $s=0.01$. The reported results in the tables indicate the general usefulness of matrix elements for the investigation of the entire spectrum of the spiked harmonic oscillator Hamiltonian for $\lambda>0$ and arbitrary $\alpha$ in any dimensions and for any angular momentum number $l$. It is also clear that we don't need a very large basis set to produce accurate bounds. It can be seen from the tables that the rate of convergence is fast for moderate values of the coupling constant $\lambda$, while for very small values of the coupling constant the rate of convergence is much slower. In general, however, throughout the whole range of values of $\lambda$, the result from the introduced basis always gives very reliable upper bounds. In summary, the basis provides a simple, uniform, and robust variational method.
\title{4.2~~Anharmonic Singular Hamiltonian}
\noindent The anharmonic singular Hamiltonians 
$$H=-{d^2\over dr^2}+{l(l+1)\over r^2}+ar^2+{b\over r^4}+{c\over r^6},\quad, a>0,c>0,\ l=0,1,2,\dots\eqno(4.12)$$ have attracted considerable attention in part because conditionally exact solutions are possible. From the mathematical point of view, this Hamiltonian is a non-trivial generalization of the spiked harmonic oscillator (4.3). Znojil
\sref{\zc-\zd} employed a Laurent series ansatz for the eigenfunctions to convert Schr\"odinger's equation into a difference equation and then used continued fraction solutions to obtain exact solutions for the ground-state and the first excited state. Kaushal and Parashar\sref{\kp} simplified Znojil's ansatz  
to obtain exact ground-state expression
$$E_0=\sqrt{a}\left(4+{b\over \sqrt{c}}\right)\quad\hbox{ subject to the constraint }\quad (2\sqrt{c}+b)^2=c(2l+1)^2+8c\sqrt{ac}.\eqno(4.13)$$
Guardiola and Ros\sref{\ah} then used a much simplifier trial wavefunction
$\psi(r)=r^{(b/\sqrt{c}+3)/2}\exp(-r^2/2-\sqrt{c}/(2r^2))$ for the case of $a=1$ and $l=0$ to obtain the exact solution for the ground state as
$$E_0=4+{b\over \sqrt{c}} \quad\hbox{ subject to the constraint condition }\quad (2\sqrt{c}+b)^2=c+8c\sqrt{c}.\eqno(4.14)$$
For example with $b=c=1$, the ground-state is $E_0=5$ and for $b=c=9$, $E_0=7$, etc. Soon afterwards, Landtman\sref{\la} performed an accurate numerical calculation and showed that for the parameters chosen by Kaushal and Parashar, although the ground-state energy they obtained agreed with the numerical calculation, their first-excited energy did not. Varshni\sref{\va}, in an attempt to resolve this problem, obtained four sets of solutions, including one constraint equation for each set and showed that the analytic expression for the energy agrees with the numerical result for any one among the ground, the first and the second excited states, depending on the particular constraint condition satisfied. For higher dimensions, by making use of certain ans\"atze for the eigenfunction, Shi-Hai Dong and Zhong-Qi Ma\sref{\sz} obtained exact closed-form solutions of (4.12) in two dimensions, where the parameters of the potentials $a, b,$ and $c$ again satisfy certain constraints. 

In order to compare our variational results with the exact eigenvalues, we have found for the exact ground-state eigenvalue $E_0=5$ of the Hamiltonian (4.12) with $l=0, a=b=c=1$, an upper bound of $E_0=5.000~006$ obtained by the diagonalization of a $14\times 14$-matrix. Further, the exact energies of $7,7,11, 11$ corresponding to $(a,b,c)=(1,9,9),(1,-7,49),(1,45, 225)$, and $(1,-24.5125,600.8623)$ respectively follow by the optimization of the matrix eigenvalues with initial guesses for the variational parameters and matrix dimensions given respectively by $(p,t,s)= (1.12,16.09,0.11), (1.03,27.47,0.07),(1.10,31.00, 0.10),$ and $(0.77, 40.82,0.01),$  and $14\times 14$   $11\times 11$,  $8\times 8$, and $7\times 7$.  These results indicate the generality and the efficiency of our approach. Note $(b,c)=(-7,49),$ and $(b,c)=(-24.5125,600.8623)$ also shows the applicability of the method in the case of $b$ negative. We further illustrate the applicability of the matrix elements to obtain accurate upper bounds to the ground-state of (4.12) for several values of $a,b$ and $c$. Indeed, for $(a,b,c)=(1,10,1), (1,10,10),$ and $(1,100,100)$, we obtain $6.679~053, 7.138~261$, and $11.791~771$ respectively which results are in excellent agreement with the exact eigenvalues  obtained by direct numerical integration of Schr\"odinger's equation. The precision of the upper bounds to any number of decimal places can be achieved by increasing $n$, the dimension of the matrix.  The energies of the excited states in arbitrary spatial dimension $d$ are similarly straightforward to find. 
\title{4.3~~Perturbed Coulomb Potentials}
\noindent Hautot\sref{\hau}, in his solutions of Dirac's equation in the presence of a magnetic field, introduced some interesting methods of solving certain second-order differential equations. One of these methods deals with the potential operator
$$V(r)= -{D/r} +Br+Ar^2, \quad A \neq 0.\eqno(4.15)$$
Hautot obtained exact solutions for only certain relations between the constants $A,B,$ and $D$. He achieved his results by applying the kinetic energy operator to an appropriate wavefunction and using the standard procedure of comparing  coefficients in the induced recurrence relations. More precisely, by introducing\sref{\hb}
$$\psi(r)=\exp\bigg(-{1\over 2}\bigg(\sqrt{A}r^2+{Br\over \sqrt{A}}\bigg)\bigg)\sum\limits_{k=0}^na_k r^{k+l}, \quad n =0,1,2,\dots$$
into the radial Schr\"odinger equation
$$\bigg({d^2\over dr^2} +{2\over r}{d\over dr} -{l(l+1)\over r^2}+E+{D\over r}-Br-Ar^2\bigg)\psi(r)=0
$$ 
One obtains the following three-term recursion relation between the coefficients $a_k$ for $(k=0,1,2,\dots)$:
$$\bigg[(k+2)(k+2l+3)\bigg]a_{k+2}+\bigg[D-{B\over \sqrt{A}}(k+2+l)\bigg]a_{k+1}+\bigg[E-\sqrt{A}(2k+2l+3)+{B^2\over 4A}\bigg] a_k=0$$
This recurrence relation terminates if $a_{k+1}=0$, that is to say
$E=E_{nl}=\sqrt{A}(2n+2l+3)-{B^2\over 4A}$
provided that the parameters $A, B,$ and $D$ satisfies the $(n+1)\times (n+1)$-determinant
$$\det\pmatrix{a_0&b_0&~&~&~&~&~&~\cr
c_1&a_1&b_1 &~&~&~&~&~\cr
~&c_2&a_2&b_2&~&~&~&~\cr
~&~&\cdot&\cdot&\cdot&\cdot&~&~\cr
~&~&~&\cdot&\cdot&c_{n-1}&a_{n-1}&b_{n-1}\cr
~&~&~&~&~&~&c_{n}&a_{n}\cr
}=0,\hbox{ 
where }
\cases{a_k=D-{B\over \sqrt{A}}(k+l+1),&~\cr\cr
	b_k=(k+1)(k+2l+2),&~\cr\cr
	c_k=E_{nl}-\sqrt{A}(2k+2l+1)+{B\over\sqrt{A}}.&~\cr}
$$
For example, we have for the ground-state energy (i.e. $k,l=0$) of 
$$V(r)=-{1\over r}+A r+(A r)^2,\quad E_0=(3+2l)A -{1\over 4},\eqno(4.16)$$
with the ground-state wavefunction given explicitly as
$\psi(r)=r^l\exp(-{1\over 2}(r+A r^2))$. This particular case was studied by Killingbeck\sref{\ka} who obtained the exact solution for the ground-state for $\beta>0$. In order to test the variational approach discussed in section 2 and 3, we have employed the matrix elements to obtain upper bounds to the exactly solvable cases such as $\beta = 0.1,1,$ and $2$, we found that the upper bounds yields $0.05,2.75,$ and $5.75$ which are in excellent agrement with the exact eigenvalues as given by (4.16). An important consequence of our variational approach are the upper bounds that are easily obtained for unconstrained values of $D, B$, and $A$. In Table 5, we have reported our variational results for $D=1$ and several values of $B$ and $A$ where we compare our results with the upper bounds obtained by the direct numerical integration of Schr\"odinger's equation\sref{\hb}.  
In arbitrary dimensions, the matrix elements discussed in Sections 2 and 3 provide a uniformly simple, straightforward, and efficient way of obtaining accurate energy bounds for the entire spectrum. In order to compare our results with those in the literature, we consider in Table 6 the radial Schr\"odinger equation in $d$-dimensions in the form 
$$-{1\over 2}\bigg({d^2\over dr^2}-{\Lambda(\Lambda+1)\over r^2}\bigg)\psi+(-{a\over r}+b r+c r^2)\psi=E\psi\eqno(4.17)$$
where $\Lambda=(d+2\ell-3)/2$.  The overall factor of ${1\over 2}$ in the kinetic-energy was incorporate in our calculations by multiplying  the kinetic energy matrix elements (2.11) by this quantity. To analyze the precision of the method, we again compare our results in Table 6 with some special cases for which the eigenvalues are known\sref{\rv}. Results for the excited states within each angular momentum subspace (labelled by $\ell$) are automatically provided for (up to the dimension of the matrix used), and arbitrary spatial dimension dimension $d$ is allowed for in the general expressions for the matrix elements. 
\title{4.4.~~The quartic double-well potential $V(r)=-\gamma r^2+ r^4,\ \gamma>0$}
\noindent The quartic double-well potential 
$$V(r)=-\gamma r^2+ r^4,\quad \gamma>0,\eqno(4.18)$$
has a long history of numerical studies (see, for example,\sref{\krt} and \sref{\weu} and the references therein). Apart from its intrinsic interest, the double-well potential also plays an important role in the quantum study of the tunnelling time problem\sref{\roy}, in spectra of molecules such as ammonia and hydrogen-bonded solids\sref{\den}. Broges et al\sref{\bd}, using supersymmetry techniques, constructed trial wave functions for variational calculations of the ground-state, first, second, and third excited-states. In their comparison with the literature, they have used the results obtained from direct numerical integration of the corresponding Schr\"odinger equation, as reported in\sref{\gj}. Unfortunately,  these numerical eigenvalues were not very accurate and the errors are higher than appear in their reported tables. In Table 7, we compare our results for the first- and third-exited states with those of Broges et al\sref{\bd}, who considered the problem in one dimension; we also include accurate numerical values.
\title{5.~~Conclusion}
We have found matrix elements for Schr\"odinger operators in $d$ spatial dimensions with spherically-symmetric potentials of the form $V(r) = \sum_q a(q)r^q.$  The matrix elements for a given angular momentum $\ell$  are calculated with respect to a finite basis $\left\{\phi_i\right\}_{i=0}^{n-1}$ comprising polynomials in $r$ with an overall factor of the form $r^{1+(t-d)/2}e^{-r^p}.$  With the inclusion of a scale parameter $s$, the upper estimates are the eigenvalues ${\cal E}^{[n]}_i(p,t,s)$ of an $n\times n$ matrix eigen equation of the form $H v = \lambda N v,$ where $N = [(\phi_i,\phi_j)].$  For best results, these estimates are optimized with respect to the three parameters $\{p,t,s\}$ for a given $n.$ For the class of problems considered, the basis has the advantage that explicit analytic expressions in terms of the Gamma function are available for all the matrix elements.  The method is robust and flexible enough to yield excellent results for the whole class of problems without the need to work with very large matrices.
\np     
\references{1}
\bigskip
   \title{Acknowledgments}
\medskip
\noindent Partial financial support of this work under Grant Nos. GP3438 and GP249507 from the 
Natural Sciences and Engineering Research Council of Canada is gratefully 
acknowledged by two of us (respectively [RLH] and [NS]).
\np
\noindent {\bf Table 1.}~~~Upper bounds $E^U$ to the ground-state eigenvalues of 
$H=-{d^2\over dr^2}+r^2+{\lambda\over r^{\alpha}}$  for different  values of $\lambda$, $\alpha$. The eigenvalues for the case $\alpha=2$ are obtain for $n=1$ and can be compared with the exact formula $2+\sqrt{1+4\lambda}$. The exponent refers to the dimension $(n)$ of the matrix used for the variational computations; the triples in parentheses refer to the approximate initial values of the parameters $\{p,t,s\}$. The small letters indicate references where the same values were obtained in the literature.
 
\bigskip
\noindent\hfil\vbox{%
\offinterlineskip
\tabskip=0pt
\halign{\tabskip=4pt
\vrule#\strut&#\strut\hfil&\vrule#
\strut&\hfil#\strut\hfil&\vrule#
\strut&\hfil#\strut\hfil&\vrule#
\strut&\hfil#\strut\hfil&\vrule#
\strut&\hfil#\strut\hfil&\vrule#
\strut&\hfil#\strut\hfil&\vrule#
\tabskip=0pt\cr
\multispan4&\multispan5&\omit&\omit\vrule\cr\noalign{\hrule\hrule}
\multispan2\vrule $\lambda$&&$\alpha=0.5$&&$\alpha=1$&&$\alpha=1.5$&&$\alpha=2$&&$\alpha=2.5$&\cr
\noalign{\hrule\hrule}
&$0.0001$&&3.000~102$^{(1,a)}$&&3.000~112$^{(1,a)}$&&3.000~138$^{(1,a)}$&&3.000~199~980&&3.000~408$^{(14,a)}$&\cr
&$~$&&$(2.00,1.00,1.00)$&&(2.00,1.00,1.00)&&(2.00,1.00,1.00)&&(2.00,1.00,1.00)&&$(2.00,1.00,0.79)$&\cr
\noalign{\hrule}
&$0.001$&&3.001~022$^{(1,d)}$&&3.001~128$^{(1,b,d)}$&&3.001~382$^{(1)}$&&3.001~998~004&&3.004~022$^{(14,c,e)}$&\cr
&$~$&&$(2.00,1.00,1.00)$&&(2.00,1.00,1.00)&&(2.00, 1.00, 1.00)&&(2.00,1.00,1.00)&&$(2.01,1.06,0.78)$&\cr
\noalign{\hrule}
&$0.01$&&3.010~226$^{(1)}$&&3.011~276$^{(1,b,d)}$&&3.013~794$^{(3)}$&&3.019~803~903&&3.036~744$^{(15,c,e)}$&\cr
&$~$&&$(2.00,1.00,1.00)$&&(1.99,1.00,0.99)&&(1.99,1.00,0.99)&&(1.99,1.01,0.99)&&$(0.75,1.56,0.01)$&\cr
\noalign{\hrule}
&$0.1$&&3.102~139$^{(3)}$&&3.112~067$^{(5,b,d)}$&&3.135~053$^{(13)}$&&3.183~215~957&&3.266~874$^{(18,c,e)}$&\cr
&$~$&&$(2.00,1.00,1.00)$&&(1.90,1.00,0.96)&&(1.59,1.00,0.47)&&(2.00,1.18,1.00)&&$(0.57,1.00,0.00)$&\cr
\noalign{\hrule}
&$1$&&3.009~204$^{(13)}$&&4.057~877$^{(14,b,d)}$&&4.141~893$^{(14)}$&&4.236~067~978&&4.317~311$^{(16,c,e)}$&\cr
&$~$&&$(2.18,1.00,0.98)$&&(0.99,1.00,0.10)&&(1.80,1.06,0.56)&&(2.00,2.23,1.00)&&$(0.69,1.09,0.009)$&\cr
\noalign{\hrule}
&$10$&&12.093~130$^{(14)}$&&10.577~483$^{(14,b,d)}$&&9.324~173$^{(14)}$&&8.403~124~237&&7.735~111$^{(6,c,e)}$&\cr
&$~$&&$(1.99,1.00,0.9)$&&(1.00,1.004,0.09)&&(2.12,1.69,1.00)&&(1.99,6.40,0.99)&&$(1.70,5.85,0.73)$&\cr
\noalign{\hrule}
}
}

\nl~~$^a$ Ref.\sref{\ke}.~~$^b$ Ref.\sref{\ac}.~~ $^c$ Ref.\sref{\aa}.
~~$^d$ Ref.\sref{\hg}.~~$^e$ Refs.\sref{\hd}, \sref{\mob}, and \sref{\pc}.

\vskip 0.5 true in
\noindent {\bf Table 2.}~~~Comparison of upper bounds for the ground state energy of the Hamiltonian $H=-{d^2\over dr^2}+r^2+{\lambda\over r^{5/2}}$ by different variational techniques. The upper bounds $E^U$ are those obtained by the present work. The exponent $(n)$ refer to the dimensions of the matrix used for the variational computations. 
\bigskip
\noindent\hfil\vbox{%
\offinterlineskip
\tabskip=0pt
\halign{\tabskip=4pt
\vrule#\strut&#\strut\hfil&\vrule#
\strut&\hfil#\strut\hfil&\vrule#
\strut&\hfil#\strut\hfil&\vrule#
\strut&\hfil#\strut\hfil&\vrule#
\strut&\hfil#\strut\hfil&\vrule#
\strut&\hfil#\strut\hfil&\vrule#
\tabskip=0pt\cr
\multispan4&\multispan5&\omit&\omit\vrule\cr\noalign{\hrule\hrule}
\multispan2\vrule $\lambda$&&Ref. \sref{\aa}&&Ref. \sref{\hd}&&Ref. \sref{\hd}&&Ref. \sref{\ha}&&$E^U$&\cr
\noalign{\hrule\hrule}
&$0.001$&&$3.004~075^{(30)}$&&3.004~074$^{(30)}$&&3.004~047$^{(5)}$&&3.004~04&&3.004~022$^{(14)}$&\cr
\noalign{\hrule}
&$0.01$&&3.039~409$^{(30)}$&&3.039~244$^{(30)}$&&3.037~474$^{(5)}$&&3.037~43&&3.036~744$^{(15)}$&\cr
\noalign{\hrule}
&$0.1$&&3.302~485$^{(30)}$&&3.296~024$^{(30)}$&&3.269~700$^{(5)}$&&3.269~28&&3.266~874$^{(18)}$&\cr
\noalign{\hrule}
&$1$&&4.329~449$^{(30)}$&&4.323~263$^{(30)}$&&4.318~963$^{(5)}$&&4.318~54&&4.317~311$^{(16)}$&\cr
\noalign{\hrule}
&$10$&&7.735~136$^{(30)}$&&7.735~114$^{(30)}$&&7.735~596$^{(5)}$&&7.735~32&&7.735~111$^{(8)}$&\cr
\noalign{\hrule}
&$100$&&17.541~890$^{(30)}$&&17.541~890$^{(30)}$&&17.542~040$^{(5)}$&&17.541~92&&17.541~890$^{(11)}$&\cr
\noalign{\hrule}
&$1000$&&44.955~485$^{(30)}$&&44.955~485$^{(30)}$&&44.955~517$^{(5)}$&&44.955~49&&44.955~485$^{(4)}$&\cr
\noalign{\hrule}
}
}
\np
\noindent {\bf Table 3.}~~~Upper bounds for the ground state energy $E^U$ for the Hamiltonain 
$H=-\Delta+r^2+{\lambda\over r^{4}}$ for several values of $\lambda$. 
We compare our upper bounds $E^U$ with the literature.  The superscript numbers are the dimension of the matrix used;
the triples in parentheses refer to the approximate initial values of the parameters $\{p,t,s\}$.

\bigskip 
\hskip 0.5 true in
\vbox{\tabskip=0pt\offinterlineskip
\def\tablerule{\noalign{\hrule}}
\def\vr{\vrule height 12pt}
\halign to350pt{\strut#\vr&#
\tabskip=1em plus2em
&\hfil#\hfil
&\vrule#
&\hfil#\hfil
&\vrule#
&\hfil#\hfil
&\vr#\tabskip=0pt\cr
\tablerule&&$\lambda$&&\bf $E^U$&&\bf $E$ &\cr\tablerule
&&0.0001&&$3.022~275^{22}$&&$3.022~275^k$&\cr
&&~&&$(0.30,3.89, 0.00)$&&$~$&\cr\tablerule
&&0.001&&$3.068~763^{20}$&&$3.068~763^{a,c},3.06877^b$&\cr
&&~&&$(0.33,8.00,0.00)$&&$~$&\cr\tablerule
&&0.005&&$3.148~352^{20}$&&$3.148352^{c},3.14839^h,3.14835^g$&\cr
&&~&&$(0.44,2.36, 0.001)$&&$3.14664^d,3.148352^d,3.05319^n$&\cr
&&~&&$~$&&$3.14835^o$&\cr\tablerule
&&0.01&&$3.205~069^{20}$&&$3.205067^{a},3.20508^b,3.20442^d$&\cr
&&~&&$(0.44,2.10, 0.00)$&&$3.20507^k,3.20527^h, 3.20507^g$&\cr
&&~&&$~$&&$3.205067^d,3.07522^n,3.23775^k$&\cr
&&~&&$~$&&$3.20548^m, 3.20507^o,3.24980^p$&\cr\tablerule
&&0.1&&$3.575~559^{14}$&&$3.575552^{a,c},3.57557^b$&\cr
&&~&&$(0.42,9.30, 0.00)$&&$3.57555^k,3.62644^k,3.60044^p$&\cr\tablerule
&&0.4&&$4.031~971^{22}$&&$4.031971^i,4.031971^f$&\cr
&&~&&$(0.50,3.89, 0.00)$&&$~$&\cr\tablerule
&&1&&$4.494~179^{11}$&&$4.494178^{a},4.49418^b$&\cr
&&~&&$(0.70,11.00,0.00)$&&$4.49418^k,4.54879^k$&\cr\tablerule
&&10&&$6.606~625^{14}$&&$6.606623^{a,c},6.60662^b,6.60662^k$&\cr
&&~&&$(0.44,18.00, 0.00)$&&$6.64978^k,6.609~66^p$&\cr\tablerule
&&100&&$11.265~080^{7}$&&$11.265080^{a},11.26508^b,11.26508^{k}$&\cr
&&~&&$(0.49,72.00, 0.00)$&&$11.265~86^p$&\cr\tablerule
&&1000&&$21.369~464^6$&&$21.369463^{a,c,l}, 21.36946^b$&\cr
&&~&&$(0.62,100.80, 0.00)$&&$21.370~26^p$&\cr\tablerule
}}
\medskip

$$\matrix{
^a\ \hbox{Ref.}\sref{\ro}.&
^b\ \hbox{Ref.}\sref{\ad}.&
^c\ \hbox{Ref.}\sref{\ro}.&
^d\ \hbox{Ref.}\sref{\sol}.&
^e\ \hbox{Ref.}\sref{\ke}.&
^f\ \hbox{Ref.}\sref{\pc}.\cr
^g\ \hbox{Ref.}\sref{\kc}.&
^h\ \hbox{Ref.}\sref{\detw}.&
^i\ \hbox{Ref.}\sref{\zd}.&
^j\ \hbox{Ref.}\sref{\fer}.&
^k\ \hbox{Ref.}\sref{\pc}.&
^l\ \hbox{Ref.}\sref{\kc}.\cr
^m\ \hbox{Ref.}\sref{\detw}.&
^n\ \hbox{Ref.}\sref{\zd}.&
^o\ \hbox{Ref.}\sref{\hg}.&
^p\ \hbox{Ref.}\sref{\harr}.&
^q\ \hbox{Ref.}\sref{\kola}.&
^s\ \hbox{Ref.}\sref{\ha}.\cr
}$$

\np
\noindent {\bf Table 4.}~~~A comparison between the upper bounds for the Hamiltonian $H=-{d^2\over dr^2}+r^2+{A\over r^2}+{\lambda\over r^4}$, for a wide range of values of $A=l(l+1)$ and $\lambda$, using the present work $E^U$ and the bounds $E_a^U$ obtained by Aguilera-Navarro et al\sref{\ag} (see also \sref{\hm} and \sref{\ro}). Accurate numerical results $E$ obtained by direct numerical solution of Sch\"odinger's equation are also presented. The exponent refers to the dimension $(n)$ of the matrix used for the variational computations; the triples in parentheses refer to the approximate initial values of the parameters $\{p,t,s\}$.
\bigskip
\bigskip
\vbox{\tabskip=0pt\offinterlineskip
\def\tablerule{\noalign{\hrule}}
\def\vr{\vrule height 12pt}
\halign to470pt{\strut#\vr&#
\tabskip=1em plus2em
&\hfil#\hfil
&\vrule#
&\hfil#\hfil
&\vrule#
&\hfil#\hfil
&\vrule#
&\hfil#\hfil
&\vrule#
&\hfil#\hfil
&\vr#\tabskip=0pt\cr
\tablerule&&$\lambda$&&$l$&&$E_a^U$&&$E^{U}$&&$E$&\cr\tablerule
&&$0.001$&&$3$&&$~9.000~114~279$&&$9.000~114~279^{(11)}~~(1.99,7.00,0.99)$&&$9.000~114~279$&\cr
&&~&&$4$&&$11.000~063~490$&&$11.000~063~490^{(11)}~~( 1.90,9.00,1.03)$&&$11.000~063~490$&\cr
&&~&&$5$&&$13.000~040~403$&&$13.000~040~403^{(11)}~~ (1.98,9.00,0.96)$&&$13.000~040~403$&\cr\tablerule
&&$0.01$&&$3$&&$~9.001~142~268$&&$9.001~142~199^{(11)}~~(2.10,7.22, 0.98)$&&$9.001~142~199$&\cr
&&~&&$4$&&$11.000~634~795$&&$11.000~634~788^{(11)}~~(2.04,6.91, 1.00)$&&$11.000~634~788$&\cr
&&~&&$5$&&$13.000~404~001$&&$13.000~404~000^{(11)}~~(2.01, 7.00, 1.07)$&&$13.000~404~000$&\cr\tablerule
&&$0.1$&&$3$&&$~9.011~370~328$&&$9.011~364~024^{(13)}~~(2.00,5.00,1.19)$&&$9.011~364~024^{*}$&\cr
&&~&&$4$&&$11.006~336~739$&&$11.006~336~013^{(13)}~~ (2.00,3.99,0.84)$&&$11.006~336~013^*$&\cr
&&~&&$5$&&$13.004~036~546$&&$13.004~036~433^{(8)}~~(1.85,5.97,0.79)$&&$13..004~036~433$&\cr\tablerule
&&$1$&&$3$&&$~9.109~013~250~38$&&$9.108~657~991^{(14)}~~(1.70,4.00,0.61)$&&$9.108~657~991^*$&\cr
&&~&&$4$&&$11.062~293~143~4$&&$11.062~241~722^{(11)}~~(1.90,8.99,0.78)$&&$11.062~241~719^*$&\cr
&&~&&$5$&&$13.040~025~483~8$&&$13.040~015~183^{(8)}~~(1.81,7.19, 0.77)$&&$13.040~015~183$&\cr
\tablerule
}}
\bigskip\medskip
\noindent{\bf Table 5.}~~~Upper bounds for the Hamiltonian $H=-{d^2\over dr^2}-{D\over r}+Br+Ar^2$ for different values of the parameters $B$ and $A$. The numerical results in the brackets are the exact eigenvalues as obtained by direct numerical integration of Schr\"odinger equation. The triples in parentheses refer to the approximate initial values of the parameters $\{p,t,s\}$.
\medskip
\hskip 0.2 true in
\vbox{\tabskip=0pt\offinterlineskip
\def\tablerule{\noalign{\hrule}}
\def\vr{\vrule height 12pt}
\halign to400pt{\strut#\vr&#
\tabskip=1em plus2em
&\hfil#\hfil
&\vrule#
&\hfil#\hfil
&\vrule#
&\hfil#\hfil
&\vrule#
&\hfil#\hfil
&\vr#\tabskip=0pt\cr
\tablerule&&$D$&&$B$&&$A$&&$E^U$&\cr\tablerule
&&1&&$1$&&$2$&&$3.656~525\quad (3.657)$&\cr
&&~&&$~$&&$~$&&$8\times 8,(1.99,1.00,0.73)$&\cr\tablerule
&&1&&$0.1$&&$1$&&$1.885~424\quad (1.885)$&\cr
&&~&&$~$&&$~$&&$11\times 11,(1.92,1.00,0.75)$&\cr\tablerule
&&1&&$0.5$&&$1$&&$2.277~581\quad (2.278)$&\cr
&&~&&$~$&&$~$&&$10\times 10,(2.04,1.00,0.86)$&\cr\tablerule
&&1&&$0.1$&&$0.1$&&$0.378~305 \quad (0.378)$&\cr
&&~&&$~$&&$~$&&$12\times 12,(2.21,1.00,1.63)$&\cr\tablerule
&&1&&$0.01$&&$1$&&$1.795~268\quad (1.795)$ &\cr
&&~&&$~$&&$~$&&$8\times 8,(1.08,1.00,0.09)$&\cr\tablerule
&&1&&$0.001$&&$1$&&$1.786~212\quad (1.786)$&\cr
&&~&&$~$&&$~$&&$8\times 8,(2.04,1.00,0.93)$&\cr\tablerule
}}

\noindent {\bf Table 6.}~~~Comparison of the eigenvalues for $H=-{d^2\over dr^2}-{D\over r}+Br+Ar^2$ for different values of $D$, $B$, and $A$ where $E^N$ is calculated from the shifted $1/N$ expansion\sref{\rv}, the exact supersymmetric values $E^s$\sref{\rv} and the upper bounds $E^U$ obtained by the method of the present paper (diagonalization of the $d\times d$ matrix elements then minimizing with repect to the parameters $\{p,t,s\}$). 
\bigskip
\noindent\hfil\vbox{%
\offinterlineskip
\tabskip=0pt
\halign{\tabskip=5pt
\vrule#\strut&#\strut\hfil&\vrule#\strut&\hfil#\strut\hfil&\vrule#\strut&\hfil#\strut\hfil&\vrule#\strut&\hfil#\strut\hfil&\vrule#\strut&\hfil#\strut\hfil&\vrule#\strut&\hfil#\strut\hfil&\vrule#\strut&\hfil#\strut\hfil&\vrule#\strut\tabskip=0pt\cr
\multispan2&\multispan4{\hrulefill}&\multispan4{\hrulefill}&\multispan4\hrulefill\cr
\multispan5&\multispan5&\omit&\omit\vrule\cr\noalign{\hrule}
\multispan2\vrule $l$&&$D$&&
		$B$&&$A$&&$E^N$&&$E^s$&&$E^U$&\cr
\noalign{\hrule}
&$0$&&1&&0.447~21&&0.1&&0.171~66&&0.170~82&&$0.170~82^6$&\cr
\noalign{\hrule}
&$1$&&1&&0.223~61&&0.1&&0.993~37&&0.993~03&&$0.993~04^8$&\cr
\noalign{\hrule}
&$2$&&1&&0.149~07&&0.1&&1.509~79&&1.509~69&&$1.509~69^3$&\cr
\noalign{\hrule}
&$3$&&1&&0.111~80&&0.1&&1.981~24&&1.981~21&&$1.981~21^3$&\cr
\noalign{\hrule}
&$0$&&1&&1.414~21&&1.0&&1.627~56&&1.621~32&&$1.621~32^4$&\cr
\noalign{\hrule}
&$1$&&1&&0.707~11&&1.0&&3.411~41&&3.410~53&&$4.410~54^8$&\cr
\noalign{\hrule}
&$2$&&1&&0.471~40&&1.0&&4.894~40&&4.894~19&&$4.894~19^4$&\cr
\noalign{\hrule}
&$3$&&1&&0.353~55&&1.0&&6.332~78&&6.332~71&&$6.332~71^1$&\cr
\noalign{\hrule}
&$0$&&1&&4.47214&&10&&6.226~80&&6.208~20&&$6.208~22^4$&\cr
\noalign{\hrule}
&$1$&&1&&2.23607&&10&&11.057~19&&11.055~34&&$11.055~34^2$&\cr
\noalign{\hrule}
&$2$&&1&&1.49071&&10&&15.59732&&15.59692&&$15.59692^4$&\cr
\noalign{\hrule}
&$3$&&1&&1.11803&&10&&20.093~49&&20.093~36&&$20.093~37^{10}$&\cr
\noalign{\hrule}
&$0$&&1&&14.14214&&100&&20.753~21&&20.713~20&&$20.713~20^4$&\cr
\noalign{\hrule}
&$1$&&1&&7.07107&&100&&35.233~90&&35.230~34&&$35.230~34^5$&\cr
\noalign{\hrule}
&$2$&&1&&4.71405&&100&&49.442~67&&49.441~92&&$49.441~92^9$&\cr
\noalign{\hrule}
&$3$&&1&&3.535~53&&100&&63.608~60&&63.608~36&&$63.608~37^{8}$&\cr
\noalign{\hrule}
&$0$&&1&&44.721~36&&1000&&66.65904&&66.58204&&$66.582~04^3$&\cr
\noalign{\hrule}
&$1$&&1&&22.360~68&&1000&&111.685~01&&111.678~40&&$111.678~40^7$&\cr
\noalign{\hrule}
&$2$&&1&&14.907~12&&1000&&156.470~58&&156.469~20&&$156.469~20^7$&\cr
\noalign{\hrule}
&$3$&&1&&11.180~34&&1000&&201.215~30&&201.214~87&&$201.214~87^{7}$&\cr
\noalign{\hrule}
}
}
\medskip

\noindent{\bf Table 7.}~~~Upper bounds for the Hamiltonian $H=-{d^2\over dr^2}-\gamma r^2+r^4$ with different values of $\gamma.$ $E_1(V)$ and $E_3(V)$ represent the values obtained from the variational method discussed by Broges et al, and $E_1^U$ and $E_3^U$ are from the present work (with a $10\times 10$-matrix). We have also included accurate numerical results $E_1^N$ and $E_3^N$ obtained by direct numerical integration of Schr\"odinger's equation. 

\bigskip
\noindent\hfil\vbox{%
\offinterlineskip
\tabskip=0pt
\halign{\tabskip=5pt
\vrule#\strut&#\strut\hfil&\vrule#\strut&\hfil#\strut\hfil&\vrule#\strut&\hfil#\strut\hfil&\vrule#\strut&\hfil#\strut\hfil&\vrule#\strut&\hfil#\strut\hfil&\vrule#\strut&\hfil#\strut\hfil&\vrule#\strut&\hfil#\strut\hfil&\vrule#\strut\tabskip=0pt\cr
\multispan2&\multispan4{\hrulefill}&\multispan4{\hrulefill}&\multispan4\hrulefill\cr
\multispan5&\multispan5&\omit&\omit\vrule\cr\noalign{\hrule}
\multispan2\vrule $\gamma$&&$E_1(V)$&&
		$E_1^N$&&$E_1^U$&&$E_3(V)$&&$E_3^N$&&$E_3^U$&\cr
\noalign{\hrule}
&$0.1$&&3.710~64&&3.708~93&&3.708~93&&11.542~58&&11.488~48&&$11.488~48$&\cr
\noalign{\hrule}
&$0.2$&&3.618~90&&3.617~01&&3.617~01&&11.386~92&&$11.331~27$&&$11.331~27$&\cr
\noalign{\hrule}
&$0.3$&&3.525~96&&3.523~87&&3.523~87&&11.230~45&&11.173~10&&$11.173~10$&\cr
\noalign{\hrule}
&$0.4$&&3.431~79&&3.429~47&&3.429~47&&11.073~07&&11.013~97&&$11.013~97$&\cr
\noalign{\hrule}
&$0.5$&&3.336~36&&3.333~78&&3.333~78&&10.914~77&&10.853~87&&$10.853~87$&\cr
\noalign{\hrule}
&$0.6$&&3.239~62&&3.236~76&&3.236~76&&10.755~56&&10.692~80&&$10.692~80$&\cr
\noalign{\hrule}
&$0.7$&&3.141~55&&3.138~37&&3.138~37&&11.595~47&&10.530~74&&$10.530~74$&\cr
\noalign{\hrule}
&$0.8$&&3.042~10&&3.038~56&&3.038~56&&10.434~48&&10.367~70&&$10.367~70$&\cr
\noalign{\hrule}
&$0.9$&&2.941~23&&2.937~30&&2.937~30&&10.272~58&&10.203~67&&$10.203~67$&\cr
\noalign{\hrule}
&$1.0$&&2.838~91&&2.834~54&&2.834~54&&10.109~78&&10.038~65&&$10.038~65$&\cr
\noalign{\hrule}
&$2.0$&&1.726~29&&1.713~03&&1.713~03&&8.433~95&&8.332~87&&$8.332~87$&\cr
\noalign{\hrule}
}
}
\hfil\vfil
\end

We end this section with alternative representation of the matrix elements that can be used in variational computations of the eigenvalues. We note that the radial Schr\"odinger equation for a spherically symmetric potential $V(r)$ in $d$-dimensional space
$$-{d^2\psi\over dr^2}-{d-1\over r}{d\psi\over dr}+{l(l+d-2)\over r^2}\psi+V(r)\psi=E\psi,\quad \psi(r)\in L^2([0,\infty),r^{d-1}dr)\eqno(2.12)$$ 
where $l$ denotes the angular momentum numbers, is transformed, through elimination of the first derivative term, to 
$$-{d^2R\over dr^2}+\bigg[{(d-3)(d-1)\over 4r^2}+{l(l+d-2)\over r^2}+V(r)\bigg]R=ER,\quad R(r)\in L^2([0,\infty),dr)\eqno(2.13)$$
where $R$, the reduced radial wave function, is defined by
$R(r)=r^{(d-1)/2}\psi(r)$. Through the simple deformation
$${(d-3)(d-1)\over 4}+l(l+d-2)=(m+{1\over 2}(d-3))(m+{1\over 2}(d-3)),\eqno(2.14)$$
Eq.(2.13) can be written as 
$$-{d^2R\over dr^2}+\bigg[{(m+{1\over 2}(d-1))(m+{1\over 2}(d-3))\over r^2}+V(r)\bigg]R=ER,\quad R(r)\in L^2([0,\infty),dr)\eqno(2.15)$$
which modified Schr\"odinger equation (2.12) in arbitrary dimensions into a simple analogy of the 3-dimesnional radial Schr\"odinger equation.  Herein, for given $l$ and $d$ in (2.12), we should use the correspondence $m$ from (2.14).
We note further that in (2.15), $d$ and $l$ enter into the expression (2.15) in the form of the combination $2l+d$. Consequently, the solutions for a particular central potential $V(r)$ are the same as long as $d+2l$ remains unaltered. For examlple, the ground-state eigenfunctions $\psi$ and eigenvalues $E$ in 4-dimensional space are identical to the $p$-wave (i.e. $l'=1$) 2-dimensional solutions, the ground-stated eigenfunction $\psi$ and eigenvalue $E$ in fifth-dimensional space are identical to the $p$-wave (i.e. $l'=1$) three-dimensional solutions, etc.  In the new sitting, our trial wavefunction $\psi_i(r)=r^{g/2+i}\exp(-r^p/2)$ where $\psi(r)\in L^2([0,\infty),r^{d-1}dr)$ takes the form $R_i(r)=r^{g/2+(d-1)/2+i}\exp(-r^p/2)$ in $L^2([0,\infty),dr)$. This mapped wavefunction does not leads to changes in the normalization matrix elements $N_{ij}$, namely 
$$N_{ij}(p,g) = \left(R_i,R_j\right) =\frac{1}{p}\Gamma\left(\frac{i+j+g+d}{p}\right),$$
also the matrix elements of the potential term $r^q,$ remain unalternate in this new representation, namely,
$$P_{ij}(q,p,g)  = \left(R_i,r^qR_j\right) =\frac{1}{p}\Gamma\left(\frac{i+j+g+q+d}{p}\right),\quad g > -(q + d).$$
The only changes in the matrix elements due to the mapping $R_i(r)=r^{(d-1)/2}\psi_i(r)$ take a place in the Kinetic energy expressions which now reads
$$\eqalign{K_{ij}(p,g)&={1\over p}\bigg[\bigg((m+{1\over 2}(d-1))(m+{1\over 2}(d-3))+(i+g/2)(j+g/2)\bigg)\Gamma\left({i+j+g+d-2\over p}\right)\cr &-(i+j+g){p\over 2}\Gamma\left( {i+j+g+p+d-2}\over p \right)+{p^2\over 4}\Gamma\left ( {i+j+g+2p+d-2}\over p\right)\bigg].}\eqno(2.16)
$$ for $d$ and $m$ as given by (2.14).
\end
  \title{1.~~Introduction}
We study quantum mechanical Hamiltonians $H = -\Delta + V(r)$ in $d$ dimensions, where $V$ is a spherically-symmetric potential that supports discrete eigenvalues.  That is to say,  $r = |\mb{r}|$ and $\mb{r}\in \Re^d.$  We estimate the spectrum in an $n$-{dimensional} trial space spanned by functions of with the form
$$\Phi(r) = \sum_{i = 0}^{n-1}c_i \phi_i(r) = \sum_{i = 0}^{n-1} c_i r^{i+g/2}e^{-\half r^p}.\eqno{(1.1)}$$
\nl If the potential is chosen to be a linear combination of powers
$$V(r) = \sum_{q}a(q)r^q,\eqno{(1.2)}$$
\nl then all the matrix elements of $H$ may be expressed explicitly in terms of the Gamma function.  The expressions obtained will be functions of the parameters $g$ and $p$, and also of a scale parameter $s$ to be introduced later.  Thus we have $n + 3$ variational parameters with which to optimize upper estimates to the spectrum of $H$, with one degree of freedom being employed for normalization. 
 
Systems with Hamiltonians of this type have enjoyed wide attention in the literature of quantum mechanics\sref{\harr-\weu}. This interest arises particularly from the usefullness of these  problems as models in atomic and molecular physics.  Many numerical and analytical techniques have been used to tackle Hamiltonians of this form. If the potential is highly singular, the parameter $g$ must be chosen sufficiently large. If the most singular potential term is of the form $1/r^{\alpha},$ then we require $g > g_0 = \alpha - d.$  There is of course no need to explore $g$ variationally beyond $g_0 + 1$ since this would essentially be equivalent to changing the coefficients $\{c_i\}.$ 
In Section~2 we derive the general matrix elements and show how the minimization with respect to scale $s$ can be easily included. In section~3 we disuss some numerical issues not the least of which is the usefulness of the reduction of the matrix eigen equations to symmetric form by first diagonalizing the `normalization' matrix $N = [(\phi_i,\phi_j)].$  The dependence of the eigenvalues on the parameters $\{p,g,s\}$ may be rather complicated.  Since changes to scale $s$ do not involve the recomputation of the basic matrix elements, a policy which emerges is to fix $n$, always optimize fully with respect to scale $s,$ and, if necessary, to optimize only roughly with respect to $g$ and $p$by exploring a few values; if higher accuracy is required, $n$ may then be increased. In Section~4 the matrix elements are applied to a variety of problems and the results are compared with those found in earlier work.   
 We suppose that the  Hamiltonian operators in this paper have domains ${\cal D}(H)\subset L^2(\Re^d),$ they are bounded below, essentially self adjoint, and have at least one discrete eigenvalue at the bottom of the spectrum.  This, of course, implies that the potential cannot be dominated by repulsive terms. Because the potentials are spherically symmetric, the discrete eigenvalues $E_{n\ell}^d$ can be labelled by two quantum numbers, the total angular momentum $\ell = 0,1,2,\dots,$ and a `radial' quantum number, $n = 1,2,3,\dots,$ which counts the eigenvalues in each angular-momentum subspace. These eigenvalues satisfy the relation $E^d_{n\ell}\le E^d_{m\ell},\ n<m.$ With our labelling convention, the eigenvalue $E^d_{n\ell}(q)$ in $d\geq 2$ spatial dimensions has degeneracy $1$ for $\ell=0$ and, for $\ell>0,$ the degeneracy is given\sref{\movr} by the function $\Lambda(d,\ell)$, where
$$\Lambda(d,\ell)=(2\ell+d-2)(\ell+d-3)!/\{\ell!(d-2)!\},\quad d \geq 2, \ \ell>0.\eqno{(1.3)} $$
Many techniques have been applied to approximate the spectrum of singular potentials of the form (1.2) using perturbation, variational, and geometrical approximation techniques\sref{\harr-\weu}. Exact solutions for the energy may be obtained in some special cases by first choosing a wave function with parameters, and then finding a potential of the form (1.2) for which this wave function is an eigenfunction; this is possible only when certain constraints are satisfied between the parameters $\{a(q)\}$ as we shall discuss later. 
  \title{2.~~Matrix elements}   
The matrix elements we seek are 
$$H_{ij}=\left(\Phi_i,-\Delta\Phi_j\right)+ \sum_q a(q)\left(\Phi_i, r^q \Phi_j\right),\eqno{(2.1)} $$ 
\nl where $\Phi = \phi Y_{\ell}$ and $Y_{\ell}(\theta_1,\theta_2, \dots,\theta_{d-1})$ is a normalized generalized spherical harmonic. For each potential term $r^q,$ if we everywhere omit the constant angular factor (equal to $4\pi$ in the case $d = 3$), we find 
$$\eqalign{P_{ij}(q,p,g) &= \left(\Phi_i,r^q\Phi_j\right) = \int_0^{\infty}r^{i+j+g+q+d-1}e^{-r^p}dr,\quad i,j = 0,1,2,\dots,\cr & = \frac{1}{p}\Gamma\left(\frac{i+j+g+q+d}{p}\right),\quad g > -(q + d).}\eqno{(2.2)}$$
\nl This type of integral is found by setting $x = r^p,$ and using the differential relation $r^kdr = (1/p)x^{(k+1-p)/p} dx$  and the definition of the Gamma function.  The normalization integrals are special cases of (2.2), namely
$$N_{ij}(p,g) = \left(\Phi_i,\Phi_j\right) = P_{ij}(0,p,g)= \frac{1}{p}\Gamma\left(\frac{i+j+g+d}{p}\right).\eqno{(2.3)}$$
\nl For the kinetic energy we first need the expression for the Laplacian in $d$ dimensions\sref{\som}.  More specifically, we need its effect on a wave function $\Phi = \phi Y_{\ell}$ with a spherically-symmetric factor $\phi(r)$ and a generalized spherical harmonic factor $Y_{\ell}.$ If we remove the spherical harmonic factor after the action of the Laplacian on $\Phi$ we obtain\sref{\som} 
$$\frac{\Delta \Phi} {Y_{\ell}} = \phi''(r) +\frac{d-1}{r}\phi'(r) - \frac{\ell(\ell+d-2)}{r^2}\phi(r).\eqno{(2.4)}$$
\nl We now suppose that $\Psi = \psi Y_{\ell}$ and that the spherical harmonic factor is normalized.  We compute the general matrix element $(\Psi, -\Delta \Phi)$ and perform an integration by parts, with the assumptions that $\psi$ and $\phi$ are twice differentiable and in $L^2\left([0,\infty),r^{d-1}dr\right),$ and vanish at $r = 0$, to find that the second term in (2.4) cancels with a term in the integration by parts and we have:
$$(\Psi, -\Delta \Phi) = \int_0^{\infty}\left(\psi'(r)\phi'(r) + \frac{\ell(\ell+d-2)}{r^2}\psi(r)\phi(r)\right)r^{d-1}dr.\eqno{(2.5)}$$
\nl The assumption of $\psi(0) = \phi(0) = 0$ has the advantage of a single general formula, but the disadvantage that we cannot at the same time accommodate the even parity solutions in dimension $d = 1.$ Finally, by applying (2.5) in the case that $\psi = \phi_i$ and $\phi = \phi_j$ we find
the general kinetic energy matrix element to be given by
$$K_{ij}(p,g)={1\over p}\left[(\ell(\ell+d-2)+(i+g/2)(j+g/2))\Gamma\left({{i+j+g+d-2}\over p}\right)\right.$$ 
$$\left.-(i+j+g){p\over 2}\Gamma\left({{i+j+g+p+d-2}\over p} \right)+{{p^2}\over 4}\Gamma\left( {{i+j+g+2p+d-2}\over p}\right)\right].\eqno{(2.6)} $$
\nl These formulas may be used as they stand for all dimensions $d \geq 3$ provided that $g$ is chosen sufficiently large $g > -(d+\hat{q})$ to control the most singular potential term $r^{\hat{q}}.$  For dimensions $d = 2$ and $d = 1$ special considerations may be required. For example, for dimension $d = 2$ we observe that if $\ell > 0,$ then it is necessary that $g > 0$ even when there are no singular potential terms.  In dimension $d = 1,$ we always require $\ell = 0$ but note that only the eigenvalues corresponding to the odd states are allowed for in our formulation; moreover, for non-singular potentials, for example, if $g$ and $p$ are small, all but the last term in the kinetic energy (2.6) must be set to zero by means of limits.\medskip 
   
We now consider the problem of minimizing $(\Phi,H\Phi)$ with respect to the vector $v$ of coefficient $\{ c_i\}_{i=0}^{n-1}$ subject to the constraint that $(\Phi,\Phi)=1$. We assume here that each term in the wavefunction sum has the spherical-harmonic factor $Y_{\ell}$ with the same value of $\ell$. This leads to the following matrix eigenvalue problem
$$Hv= {\cal E} N v \eqno{(2.7)}$$
By the min-max characterization of the spectrum\sref{\reed}, the eigenvalues of this matrix equation are upper bounds to the unknown exact eigenvalues $E_{i\ell}, i=1,2,\dots n-1.$ We assume that these discrete eigenvalues of $H$ are either known to exist, or indeed are demonstrated to exist by the results of this variational estimate.
\nl By considering scaled radial wave functions of the form 
$$\phi_s(r)=\phi(r/s),\eqno{(2.8)} $$ 
we find that factors of $s$ remain only according to the dimensions of the terms. In effect, when using the scaled wave functions (8), we can leave the matrix $N$ unchanged and replace the matrix for $H$ by
$$H_{ij}(s)={1\over s^2}K_{ij}(p,g)+\sum_q a(q)s^q P_{ij}(p,g).\eqno{(2.9)} $$
Thus the upper bounds we seek are provided by the eigenvalues of the matrix equation
$$H(s)v = {\cal E} N v,\eqno{(2.10)}$$
 which now depend, for a given $n$ and $\ell,$ on $s, p$ and $g$ and we write
$$E_{i\ell} \leq {\cal E}_{i\ell}={\cal E}_{i\ell}(p,g,s),\quad i = 0,1,2 \dots n-1.\eqno(2.11)$$
\nl The problem now is to find these upper estimates and minimize them with respect to the three parameters $\{p,g,s\}$.

We end this section with a brief note concerning a commonly-used alternative representation for the problem we are studying:  the main point is that the matrix elements obtained from this form are identical to those we have obtained above.  The radial Schr\"odinger equation for a spherically symmetric potential $V(r)$ in $d$-dimensional space is given by
$$-{d^2\psi\over dr^2}-{d-1\over r}{d\psi\over dr}+{l(l+d-2)\over r^2}\psi+V(r)\psi=E\psi,\quad \psi(r)\in L^2([0,\infty),r^{d-1}dr)\eqno(2.12)$$ 
where $l$ denotes the angular momentum quantum number; in the case $d = 1,$ we always have $\ell = 0,$ otherwise $\ell = 0,1,2,\dots.$  A correspondence to a problem in one dimension is obtained with the aid of a radial wave-function $R(r)$ expressed in the form
$$R(r)=r^{(d-1)/2}\psi(r),\quad R(0) = 0.\eqno{(2.13)}$$
\nl If we now re-write (2.12) in terms of this new radial function, we obtain the following Schr\"odinger equation for a problem on the half line with a Dirichlet boundary condition at $r = 0$
$$-{d^2R\over {dr^2}}+ UR=ER,\quad R\in L^2([0,\infty),dr),\eqno(2.14)$$
\nl where the effective potential $U(r)$ is given by
$$U(r) = V(r) + {{(2\ell+d-1)(2\ell+d -3)}\over {4  r^2}}.\eqno{(2.15)}$$
\nl We note that in (2.15), $d$ and $l$ enter into the expression for $U(r)$ in the combination $2l+d$: consequently, the solutions for a given central potential $V(r)$ are the same provided $d+2l$ remains unaltered.  In the new setting, our new trial wave functions $R_i(r) = r^{(d-1)/2}\phi_i(r),$ where $\phi_i(r)\in L^2([0,\infty),r^{d-1}dr)$, take the explicit form 
$$R_i(r)=r^{g/2+(d-1)/2+i}\exp(-r^p/2) \in L^2([0,\infty),dr).\eqno{(2.16)}$$
\nl This formulation of the problem leads to the same matrix elements as have we found above.  A computer algebra system might generate different-looking expressions which can then be shown to be mathematically identical. 

\noindent{\bf Table 7.}~~~Upper bounds for the Hamiltonian $H=-{d^2\over dr^2}-\gamma r^2+r^4$ with different values of $\gamma.$ $E_1(V)$ and $E_3(V)$ represent the values obtained from the variational method discussed by Broges et al, and $E_1^U$ and $E_3^U$ are from the present work (with a $10\times 10$-matrix). We have also included the numerical results $E_1^N$ and $E_3^N$ obtained by direct numerical integration of Schr\"odinger's equation. 

\bigskip
\noindent\hfil\vbox{%
\offinterlineskip
\tabskip=0pt
\halign{\tabskip=5pt
\vrule#\strut&#\strut\hfil&\vrule#\strut&\hfil#\strut\hfil&\vrule#\strut&\hfil#\strut\hfil&\vrule#\strut&\hfil#\strut\hfil&\vrule#\strut&\hfil#\strut\hfil&\vrule#\strut&\hfil#\strut\hfil&\vrule#\strut&\hfil#\strut\hfil&\vrule#\strut\tabskip=0pt\cr
\multispan2&\multispan4{\hrulefill}&\multispan4{\hrulefill}&\multispan4\hrulefill\cr
\multispan5&\multispan5&\omit&\omit\vrule\cr\noalign{\hrule}
\multispan2\vrule $\gamma$&&$E_1(V)$&&
		$E_1^N$&&$E_1^U$&&$E_3(V)$&&$E_3^N$&&$E_3^U$&\cr
\noalign{\hrule}
&$0.1$&&3.710~64&&3.708~97&&3.708~93&&11.542~58&&11.488~57&&$11.488~48$&\cr
\noalign{\hrule}
&$0.2$&&3.618~90&&3.617~04&&3.617~00&&11.386~92&&$11.331~36$&&$11.331~27$&\cr
\noalign{\hrule}
&$0.3$&&3.525~96&&3.523~90&&3.523~87&&11.230~45&&11.173~19&&$11.173~10$&\cr
\noalign{\hrule}
&$0.4$&&3.431~79&&3.429~50&&3.429~47&&11.073~07&&11.014~06&&$11.013~97$&\cr
\noalign{\hrule}
&$0.5$&&3.336~36&&3.333~81&&3.333~78&&10.914~77&&10.853~96&&$10.853~87$&\cr
\noalign{\hrule}
&$0.6$&&3.239~62&&3.236~79&&3.236~76&&10.755~56&&10.692~88&&$10.692~80$&\cr
\noalign{\hrule}
&$0.7$&&3.141~55&&3.138~40&&3.138~37&&11.595~47&&10.530~83&&$10.530~74$&\cr
\noalign{\hrule}
&$0.8$&&3.042~10&&3.038~59&&3.038~56&&10.434~48&&10.367~78&&$10.367~70$&\cr
\noalign{\hrule}
&$0.9$&&2.941~23&&2.937~33&&2.937~30&&10.272~58&&10.203~75&&$10.203~67$&\cr
\noalign{\hrule}
&$1.0$&&2.838~91&&2.834~56&&2.834~54&&10.109~78&&10.038~72&&$10.038~65$&\cr
\noalign{\hrule}
&$2.0$&&1.726~29&&1.713~04&&1.713~03&&8.433~95&&8.332~93&&$8.332~87$&\cr
\noalign{\hrule}
}
}